\documentclass[12pt,twoside]{article}
\usepackage{amssymb}
\usepackage{amsmath}
\usepackage{mathrsfs} 
\usepackage[dvips]{graphicx}
\begin{document}
\newcommand{\cA}{{\cal A}}
\newcommand{\cB}{{\cal B}}
\newcommand{\cC}{{\cal C}}
\newcommand{\cD}{{\cal D}}
\newcommand{\cE}{{\cal E}}
\newcommand{\cF}{{\cal F}}
\newcommand{\cG}{{\cal G}}
\newcommand{\cH}{{\cal H}}
\newcommand{\cI}{{\cal I}}
\newcommand{\cJ}{{\cal J}}
\newcommand{\cK}{{\cal K}}
\newcommand{\cL}{{\cal L}}
\newcommand{\cM}{{\cal M}}
\newcommand{\cN}{{\cal N}}
\newcommand{\cO}{{\cal O}}
\newcommand{\cP}{{\cal P}}
\newcommand{\cQ}{{\cal Q}}
\newcommand{\cR}{{\cal R}}
\newcommand{\cS}{{\cal S}}
\newcommand{\cT}{{\cal T}}
\newcommand{\cU}{{\cal U}}
\newcommand{\cV}{{\cal V}}
\newcommand{\cX}{{\cal X}}
\newcommand{\cW}{{\cal W}}
\newcommand{\cY}{{\cal Y}}
\newcommand{\cZ}{{\cal Z}}

\def\cl{\centerline}
\def\bd{\begin{description}}
\def\be{\begin{enumerate}}
\def\ben{\begin{equation}}
\def\benn{\begin{equation*}}
\def\een{\end{equation}}
\def\eenn{\end{equation*}}
\def\benr{\begin{eqnarray}}
\def\eenr{\end{eqnarray}}
\def\benrr{\begin{eqnarray*}}
\def\eenrr{\end{eqnarray*}}
\def\ed{\end{description}}
\def\ee{\end{enumerate}} 
\def\al{\alpha}
\def\b{\beta}
\def\bR{\bar\R}
\def\bc{\begin{center}}
\def\ec{\end{center}}
\def\d{\dot}
\def\D{\Delta}
\def\del{\delta}
\def\ep{\epsilon}
\def\g{\gamma}
\def\G{\Gamma}
\def\h{\hat}
\def\iny{\infty}
\def\La{\Longrightarrow}
\def\la{\lambda}
\def\m{\mu}
\def\n{\nu}
\def\noi{\noindent}
\def\Om{\Omega}
\def\om{\omega}
\def\p{\psi}
\def\pr{\prime}
\def\r{\ref}
\def\R{{\bf R}}
\def\ra{\rightarrow}
\def\s{\sum_{i=1}^n}
\def\si{\sigma}
\def\Si{\Sigma}
\def\t{\tau}
\def\th{\theta}
\def\Th{\Theta}
\def\vep{\varepsilon}
\def\vp{\varphi}
\def\pa{\partial}
\def\un{\underline}
\def\ov{\overline}
\def\fr{\frac}
\def\sq{\sqrt}

\def\WW{\begin{stack}{\circle \\ W}\end{stack}}
\def\ww{\begin{stack}{\circle \\ w}\end{stack}}
\def\st{\stackrel}
\def\Ra{\Rightarrow}
\def\R{{\mathbb R}}
\def\bi{\begin{itemize}}
\def\ei{\end{itemize}}
\def\i{\item}
\def\bt{\begin{tabular}}
\def\et{\end{tabular}}
\def\lf{\leftarrow}
\def\nn{\nonumber}
\def\va{\vartheta}
\def\wh{\widehat}
\def\vs{\vspace}
\def\Lam{\Lambda} 
\def\sm{\setminus}
\def\ba{\begin{array}}
\def\ea{\end{array}} 
\def\bd{\begin{description}}
\def\ed{\end{description}}
\def\lan{\langle}
\def\ran{\rangle}

\begin{center}
{\LARGE  On the degree conjecture for separability of  multipartite  quantum states }\\
~\\
Ali Saif M. Hassan \footnote{Electronic address: alisaif@physics.unipune.ernet.in} and  Pramod S. Joag\footnote{Electronic address: pramod@physics.unipune.ernet.in}\\
 Department of Physics, University of Pune,  Pune, India-411007. 
\end{center}

We settle the so-called degree conjecture for the separability of multipartite quantum states, which are normalized
graph Laplacians, first given by Braunstein {\it et al.} [Phys. Rev. A \textbf{73}, 012320 (2006)]. The conjecture states that a multipartite quantum state is separable if and only if the degree matrix of the graph associated with the state is equal to the degree matrix of the partial transpose of this graph. We call this statement to be the strong form of the conjecture. In its weak version, the conjecture requires only the necessity, that is, if the state is separable, the corresponding degree matrices match. We prove the strong form of the conjecture for {\it pure} multipartite quantum states, using the modified tensor product of graphs defined in [J. Phys. A: Math. Theor. \textbf{40}, 10251 (2007)], as both necessary and sufficient condition for separability. Based on this proof, we give a polynomial-time algorithm for completely factorizing any pure multipartite quantum state. By polynomial-time algorithm we mean that the execution time of this algorithm increases as a polynomial in $m,$ where $m$ is the number of parts of the quantum system. We give a counter-example to show that the conjecture fails, in general, even in its weak form, for multipartite mixed states. Finally, we prove this conjecture, in its weak form, for a class of multipartite mixed states, giving only a necessary condition for separability.\\

\vspace{.2in}

\textbf{I. INTRODUCTION} \\

The problem of detection and quantification of entanglement of multipartite quantum states is fundamental to the whole field of quantum information and in general to the physics of multicomponent quantum systems. Whereas entanglement of pure bipartite states is well understood, the classification of mixed states according to degree and character of their entanglement is still a matter of intense research [3,4]. Presently, the most successful approach for bipartite and multipartite entanglement is via PPT criterion [5], CCNR criterion [6], positive maps and entanglement witnesses [7,8], Bloch representation of quantum states [9,10], covariance matrices [11] and local uncertainty relations [12]. All these approaches lead to either necessary or sufficient (but not both) criteria for separability. Recently, a combinatorial approach to the separability problem is initiated by S. L. Braunstein, S. Ghosh and  S. Severini [13] and extended by us [2]. In this approach the density matrices are coded in terms of graphs with the idea of using the graph topology and operations on graphs for detection, classification and quantification of entanglement in the multipartite quantum states. S. L. Braunstein, S. Ghosh, T. Mansour, S. Severini, and R. C. Wilson [1] have made a conjecture, called degree conjecture, for the separability of multipartite quantum states. The conjecture states that a multipartite quantum state is separable if and only if the degree matrix of the graph associated with the state is equal to the degree matrix of the partial transpose (with respect to a subsystem, see below) of this graph. We call this to be the strong form of the conjecture. In its weak version it requires only the necessity, that is, if the state is separable the corresponding degree matrices match. We prove the strong version for a $m$-partite {\it pure} state (section II) and give a polynomial-time algorithm, for factorization of a $m$-partite pure state (section III). We show that the conjecture fails, in general, for mixed states (section IV). Finally, we prove the weak version of the conjecture for a class of multipartite mixed states (section V).\\

\textbf{II. THE SEPARABILITY CRITERION AND ITS PROOF}\\

 For the definition of a weighted graph with real or complex weights, denoted $(G,a),$ we refer the reader to ref.[2]. We denote by $V(G,a)$ and $E(G,a)$ the vertex set and the edge set respectively of a weighted graph. The degree of a vertex $v \in V(G,a)$ is denoted by $\mathfrak{d}_v$ and $\mathfrak{d}_{(G,a)}=\sum_v \mathfrak{d}_v$ is the degree sum of the graph. The degree matrix of a weighted graph is denoted by $\Delta(G, a)$ [2]. The combinatorial Laplacian of the weighted graph is denoted $L(G,a)$ and the generalized Laplacian of $(G,a)$ is denoted $Q(G,a)$ [2]. If the generalized Laplacian of a graph $(G,a)$ is positive semidefinite we can define the density matrix of the graph $(G,a)$ as $\si(G, a) = \fr{1}{\mathfrak{d}_{(G, a)}} Q(G, a).$  Conversely, given a density matrix we can assign a graph to it [2]. A real weighted graph gets assigned to a density matrix with all real elements, while a complex weighted graph is assigned to a density matrix with complex off-diagonal elements. The vertices of the graph are labeled by the elements of the standard basis in the state space of the multipartite quantum system which is used to set up the density matrix. Also, in what follows we use the definition and properties of the modified tensor product of graphs proved in ref. [2].
 
Let $(G,a)$ be the graph corresponding to a $m$-partite pure state in $d=d_1 d_2 \cdots d_m$ dimensional Hilbert space $\mathcal{H}=\mathcal{H}^{d_1}\otimes \mathcal{H}^{d_2} \otimes \cdots \otimes \mathcal{H}^{d_m}$, where $d_i$ is the dimension of  $\mathcal{H}^{d_i},\; 1\le i \le m$. Each vertex of  $(G,a)$ is labeled by an $m$ tuple $(v_1v_2\cdots v_m)$ where  $1 \le v_i \le d_i,\; i=1,2,\cdots,m$. In other words, we set up $(G,a)$ using the standard basis in $H$.  Since $(G,a)$ is the graph of a pure state, it must be a clique on some subset of $V(G,a)$, all vertices not belonging to this subset being isolated [2].
We divide the $m$ parts of the system in two nonempty disjoint subsets (partitions) whose union makes up  the whole system. We call them $s$ and $t$, where $t$ is the complement of $s$ in the set of all parts of the system. That is, $s$ and $t$ are the nonempty subsets of $\{1,2,\cdots,m\}$, $s \cup t =\{1,2,\cdots,m\}$ and $s \cap t = \phi$. This corresponds to $\mathcal{H}=\mathcal{H}^{(s)} \otimes \mathcal{H}^{(t)}, \; \mathcal{H}^{(s)}=\mathcal{H}^{d_{i_1}} \otimes \mathcal{H}^{d_{i_2}} \otimes \mathcal{H}^{d_{i_3}}\otimes \cdots \otimes \mathcal{H}^{d_{i_s}}, \; \mathcal{H}^{(t)}=\mathcal{H}^{d_{j_1}} \otimes \mathcal{H}^{d_{j_2}} \otimes \mathcal{H}^{d_{j_3}}\otimes \cdots \otimes \mathcal{H}^{d_{j_t}},\; \{ i_1 ,i_2,\cdots ,i_s \} = s,\;   \{j_1,j_2, \cdots ,j_t\}  = t.$ 
As  a result of this division, we can  divide the $m$ tuple $v=(v_1 v_2\cdots v_m)$ into  the  corresponding partitions (strings) which we  call $v_s,v_t.$ Thus $v_s= v_{i_1} v_{i_2} \cdots v_{i_s} $ and $v_t= v_{j_1} v_{j_2} \cdots v_{j_t}.$ We can label each vertex equivalently by $(v_s,v_t)$. We call $v_s$ The $s$ part and $v_t$ the $t$ part of the vertex label. For example,  consider a four partite system whose parts are labeled $1,2,3,4$. Let $s=\{1,4\} ,\; t=\{2,3\}$. Then a vertex label $(1122)$ can be written as $(12,12)$. A vertex label $(v_1v_2v_3v_4)$ becomes $(v_1 v_4,v_2v_3).$

\textbf{Partial transpose with partition  $s$.}
This is a graph operator denoted $T_s$ operating on $E(G,a)$ which we define separately for the graphs with real and complex weights.

\textbf{Definition 2.1} : Let $(G,a)$ be a graph with real weights. The operator $T_{s}$ is defined as follows. $$T_s : (G,a) \longmapsto (G^{T_s},a^{\prime}),\; (G,a)\;\ni\{(v_s,v_t),(w_s,w_t)\}\;  \longmapsto\;  \{(w_s,v_t),(v_s,w_t)\}\;\in\;(G^{T_s},a^{\prime}),$$with $$a^{\prime}(\{(w_s,v_t),(v_s,w_t)\})=a(\{(v_s,v_t),(w_s,w_t)\}),$$ that is, $ a^{\prime}(T_s e)= a(e)$. Note that, in general, $E(G,a)$ is not closed under  $T_s$.  The operator $T_t$  for the partition $t$ giving the complement of $s$ in the $m$-partite system is defined in the same way. Note that $T_s=T_t$.

\textbf{Definition 2.2} : Let $(G,a)$ be a graph with complex weights. The operator $T_{s}$ is defined as follows. $$T_s : (G,a) \longmapsto (G^{T_s},a^{\prime}),\; (G,a)\;\ni\{(v_s,v_t),(w_s,w_t)\}\;  \longmapsto\;  \{(w_s,v_t),(v_s,w_t)\}\;\in\;(G^{T_s},a^{\prime}),$$ with $$|a^{\prime}(\{(w_s,v_t),(v_s,w_t)\})| = |a(\{(v_s,v_t),(w_s,w_t)\})|,$$ that is, $ |a^{\prime}(T_s e)|=|a(e)|.$  Note that, in general, $E(G,a)$ is not closed under  $T_s$.  The operator $T_t$  for the partition $t$ giving the complement of $s$ in the $m$-partite system is defined in the same way. Note that $T_s = T_t$.

\textbf{Definition 2.3} : Partial transpose of $(G,a)$ with real or complex weights  with respect to partition $s$, denoted $(G^{T_s},a^{\prime})$ is the graph obtained by acting $T_s$ on $E(G,a)$. Note that $V(G,a)\;=\;V(G^{T_s},a^{\prime})$,  but in general $E(G,a) \ne E(G^{T_s},a^{\prime})$.

 An edge joining the vertices, whose labels have either the same $s$ part or the same $t$ part  or both are fixed points of $T_s$. We have $$ T_s \{(v_s,v_t),(v_s,w_t)\}\;=\; \{(v_s,v_t),(v_s,w_t)\},$$
with the condition on weights  automatically satisfied.
Similarly 
$$T_s(e) \;=\; T_s \{(v_s,v_t),(w_s,v_t)\} = \{(w_s,v_t),(v_s,v_t)\}\;=\; \{(v_s,v_t),(w_s,v_t)\},$$  with the condition on weights automatically satisfied. Note that  in the case of complex weighted graphs the action of $T_s$ on   $\{(v_s,v_t),(w_s,v_t)\}$ changes its orientation and hence  $a^{\prime}(T_s(e))\;=\;a^*(e).$ But by the definition of $T_s$ this edge is still invariant under $T_s$. Further,  the definition of the operator $T_s$ leaves the phase of  $a^{\prime}(T_s(e))$, $e \in E(G,a),$ as a free parameter. We shall use this freedom later in fixing the phases of the graphs corresponding to the factors in the tensor product decomposition of a density matrix.

If both $s$ and $t$ parts of two vertices of $e \in E(G,a)$ are the same then both vertices are identical and we have a loop, which is obviously preserved under $T_s$. Thus $T_s$ divides $E(G,a)$ into two partitions, one containing all fixed points of $T_s$, that is, edges with same $s$ or $t$ part and all loops, which we call $\mathcal{F}$ set and the other containing the remaining edges which we call $\mathcal{C}$ set. In other words $(G,a)$ is  the disjoint edge union of the two spanning subgraphs corresponding to  the $\mathcal{F}$ set and the $\mathcal{C}$ set.

Note that  $T_s^2$ is the identity operator $(T_s^2\;=\;1)$. Thus $T_s$ is its own inverse and is one to one and onto.

\textbf{Lemma 2.1} :\textit{ Let $(G,a)$ be a graph of a pure state in the Hilbert space $H=\mathcal{H}^{d_1}\otimes \mathcal{H}^{d_2} \otimes \cdots \otimes \mathcal{H}^{d_m}$. Let $(G^{T_s},a^{\prime})$  be  the partial transpose of $(G,a)$ with respect to a partition $s$ as defined above. Then $E(G,a)$ is  closed under $T_s$, $E(G,a)=E(G^{T_s} ,a^{\prime})$, if and only if $\Delta(G, a) = \Delta(G^{T_s}, a^{\prime})$.}

We  emphasize that the closure of $E(G,a)$ under $T_s$ means, for every $e \in E(G,a),\; T_s(e) \;= \; e' \in E(G,a) $ and $ a(e')\; =\; a(e)$ or  $ |a(e')|= |a(e)| $ as appropriate.

\textit{Proof :}
{ \it Only if part} :
We are given that $E(G,a)$ is closed under $T_s$. We divide $E(G,a)$ into two partitions $\mathcal{C}$  and  $\mathcal{F}$   as above. Note that if $E(G,a)$ is closed under $T_s$ then the sets $\mathcal{C}$  and  $\mathcal{F}$  are separately closed under $T_s$. Consider now the set of edges incident on a vertex in $V(G,a)$. The edges in this set which belong to $\mathcal{F}$ set  are not shifted by $T_s$.  Since $E(G,a)$ is closed under $T_s$  and $T_s^2 =1$,  every incident edge belonging to $\mathcal{C}$  is the image of an edge in $\mathcal{C}$ with same weight  ( or same absolute value for the weight) under the action of  $T_s$  on  $\mathcal{C}$. Thus the  degree of each vertex is preserved under the action of $T_s$ on $E(G,a)$, for both, real and complex weighted $(G,a)$ [2],  so that $ \Delta (G,a) = \Delta(G^{T_s},a^{\prime}).$

{\it If  part }:  We are given $ \Delta (G,a) = \Delta (G^{T_s},a^{\prime})$. The edges in  $ E(G,a)$ belonging to $\mathcal{F}$ remain in $E(G,a)$ under the action of $T_s$. Now suppose that the set $\mathcal{C}$  is not closed under $T_s$.  Since $T_s$  is its own inverse, the edge  $e$ for which $T_s (e) \notin E(G,a)$ cannot be the image of any other edge in $E(G,a)$  under $T_s$. Therefore, the degree of the end vertices of $e$ is changed under the action of $T_s$.  This contradicts the assumption  $ \Delta (G,a) = \Delta (G^{T_s},a^{\prime}).$\hfill $\blacksquare$\\

\textbf{Lemma 2.2} :\textit{ Let $(G,a)$ be the graph of a pure state in the Hilbert space   $\mathcal{H}=\mathcal{H}^{d_1}\otimes \mathcal{H}^{d_2} \otimes \cdots \otimes \mathcal{H}^{d_m}$ of a $m$-partite quantum system. Then  $(G,a)=(G_s,b) \boxdot (G_t,c) $ where $(G_s,b)$ is the graph of a pure state in the Hilbert space made up of $s$ factors $\mathcal{H}^{(s)}=\mathcal{H}^{d_{i_1}} \otimes \mathcal{H}^{d_{i_2}} \otimes \mathcal{H}^{d_{i_3}}\otimes \cdots \otimes \mathcal{H}^{d_{i_s}}$ and $(G_t,c)$ is the graph of a pure state in the Hilbert space made up of $t=m-s$ factors  $\mathcal{H}^{(t)}=\mathcal{H}^{d_{j_1}} \otimes \mathcal{H}^{d_{j_2}} \otimes \mathcal{H}^{d_{j_3}}\otimes \cdots \otimes \mathcal{H}^{d_{j_t}}$, if and only if the edge set $E(G,a)$ is closed under $T_s,\; s=\{i_1, i_2, i_3, \cdots, i_s\}.$ }

\textit{Proof :} {\it Only if part} : {\it Case I} : Graphs with real  weights.
 We are given \\

$(G,a)=(G_s,b) \boxdot (G_t,c) $
$$=  \cL(G_s,b) \otimes \cL \eta (G_t, c) \dotplus \cL(G_s,b) \otimes \cN(G_t,c) \dotplus  \cN(G_s,b) \otimes \cL(G_t,c)\\
   \dotplus  \Om(G_s, b) \otimes \Om(G_t,c)  \eqno{(1)}$$

Using the definition of the operators $ \cL, \eta, \cN, \Om$ [2] and that of the tensor product of graphs, we can make the following observations. The second term is  a spanning subgraph of $(G,a)$ each of whose edges has a common $t$ part and hence is fixed point of $T_s$. The  third term is a spanning subgraph of $(G,a)$ each of whose edges has a common $s$ part, so that each edge is a fixed point of $T_s$. The fourth term is a spanning subgraph of $(G,a)$ which contains only loops all of which are fixed points of $T_s$. Again, from the definition of the tensor product we see that in the first term, any edge $\{v_s,w_s\}$ in $\cL(G_s,b)$ with weight $b(\{v_s,w_s\})$ and any $\{v_t,w_t \}$ in  $\cL \eta(G_t,c)$ with weight  $c(\{v_t,w_t\})$  gives us, under the tensor product, two edges $\{(v_s,v_t),(w_s,w_t)\}$ and $\{(v_s,w_t),(w_s,v_t)\}$  with the same weight $b(\{v_s,w_s\}) c(\{v_t,w_t\})$, which are the images of each other under $T_s$. This proves that $E(G,a)$ is closed under $T_s$.

{\it Case II} : Graphs with complex weights .
 We are given \\
 
$(G,a)=(G_s,b) \boxdot (G_t,c) $
$$=  \cL(G_s,b) \otimes \cL (G_t, c) \dotplus \cL(G_s,b) \otimes \cN(G_t,c) \dotplus \cN(G_s,b) \otimes \cL(G_t,c)$$
  $$ \dotplus \{ \Om(G_s, b) \otimes \Om(G_t,c) \sqcup 2 \cN \cL (G_s,b) \otimes \cN \cL \eta (G_t,c)\}  \eqno{(2)}$$ 
Note  that the first three terms are similar to those in Eq. (1) and the arguments corresponding to these terms in the paragraph following Eq.(1) apply, except that we require $|a(T_s e)| = |a(e)| $. Again, fourth term correspond to graph with loops (and no edges) which are fixed points of $T_s$. This proves that $E(G,a)$ is closed under $T_s$.

{\it If part} : We begin by noting that the graph $(G,a)$ has the structure of a clique and isolated vertices, with $|V(G,a)| = d = d_1 d_2 \cdots d_m$  where $d_i$ is the dimension of $\mathcal{H}^{d_i},\; 1 \le i \le m$, because it is the graph of a pure state in  $\mathcal{H}=\mathcal{H}^{d_1}\otimes \mathcal{H}^{d_2} \otimes \cdots \otimes \mathcal{H}^{d_m}$. Let $(K_n,a)$ denote the clique in $(G,a)$ and $V_k(G,a) $ be the set of vertices on the clique. Let $|V_k(G,a)|= n$. We are given that $E(G,a)$ is closed under $T_s$. Note that all loops are on the clique and no loops are on the isolated vertices. Consider a vertex $(v_s,v_t)$ on $(K_n,a)$. Let $q$  denote the number of vertices in $(K_n,a)$  having the same $s$ part as $(v_s,v_t)$ and  $p$ denote the number of vertices on $(K_n,a)$ with the same $t$ part as $(v_s,v_t)$. We note that $p$ and $q$  are the same for all vertices on $(K_n,a)$, otherwise the set $\mathcal{C}$ is not closed under $T_s$. We draw $(K_n,a)$ as a lattice of $p$ rows and $q$ columns, such that all vertices in one row have common $s$ part and all vertices in one column have common $t$ part. Since $(K_n,a)$ is a complete graph, from figure 1, we see that any vertex $(v_s,v_t)$ has $(p-1)+(q-1)$  neighbors giving edges in the $\mathcal{F}$ set and $(p-1)(q-1)$ neighbors giving edges in the $\mathcal{C}$ set. Since $(K_n,a)$ is complete $(v_s,v_t)$ has $n-1$ neighbors giving $n=pq$. 

\begin{figure}[!ht]
\begin{center}
\includegraphics[width=10cm,height=10cm]{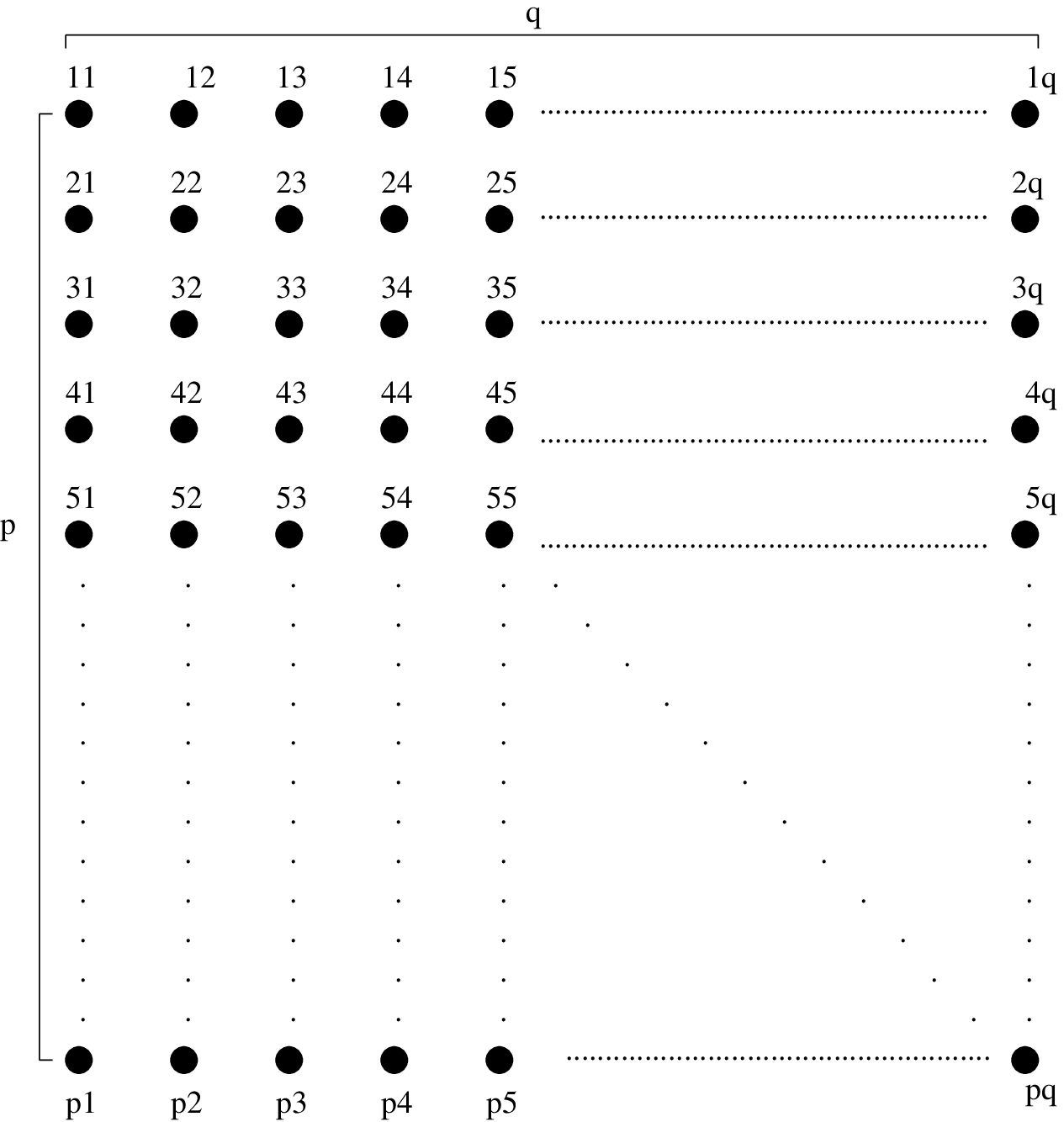}

figure 1: Every row contains vertices with same $ s$ part, and every column contains vertices with same $t$ part. In a vertex label, first number stands for the $s$ part and the second for the $t$ part. For example, the edges between the vertex 34 and all the vertices in the 3rd row and 4th column are in the set  $\mathcal{F}$ while the edges between vertex 34 and all the vertices in the four blocks obtained by deleting $3rd$ row and $4th$ column are in the set  $\mathcal{C}$.
\end{center}
\end{figure}

Now consider $\mathcal{C}$ set on $(K_n,a)$. From the definition of the tensor product of weighted graphs [2], we can factorize each pair $\{(v_s,v_t)(w_s,w_t)\}$;
$\{(w_s,v_t),(v_s,w_t)\}$  in the set $\mathcal{C}$ as the tensor product of two edges $\{v_s,w_s\}$ and $\{v_t,w_t\}$ with weights $b'$ and $c'$ satisfying  $$a(\{(v_s,v_t),(w_s,w_t)\})=  b'(\{v_s,w_s\}) \cdot c'(\{v_t,w_t\}) =a(\{(w_s,v_t),(v_s,w_t)\})$$ or $$|a(\{(v_s,v_t),(w_s,w_t)\})| =  |b'(\{v_s,w_s\})| \cdot |c'(\{v_t,w_t\})| = |a(\{(w_s,v_t),(v_s,w_t)\})|.$$  Writing each pair $\{e,T_s e\}$ in $\mathcal{C}$   set in this way and taking the disjoint edge union [2] in all these tensor products we get $(G_{\mathcal{C}},a)= (G'_1,b') \otimes (G'_2,c')$  where $(G_{\mathcal{C}},a)$ is the spanning subgraph on $(K_n,a)$ corresponding to set $\mathcal{C}$. $(G'_1,b')$ and $(G'_2,c')$ are graphs on $p$ and $q$ vertices respectively. Again, from the definition of the tensor product of weighted graphs, we know that isolated vertices in the factors produce isolated vertices in the product. Therefore we can add $(d_s-p)$ isolated vertices to $(G'_1,b')$ and $(d_t-q)$ isolated vertices to $(G'_2,c')$, where $d_s=d_{i_1} d_{i_2} \cdots d_{i_s}$ and $d_t=d_{j_1} d_{j_2} \cdots d_{j_t}$. We call this graphs $\cL(G'_s,b')$ and $\cL(G'_t,c')$  (the operator $\cL$ defined in [2] removes loops from a graph). The tensor product $\cL(G'_s,b') \otimes \cL(G'_t,c')$ gives the spanning subgraph of $(G,a)$ corresponding to set $\mathcal{C}$. Now consider a row in figure 1 containing vertices with common  $s$ part  say $v_s$. This row generates $q(q-1)/2$ edges of the form $\{(v_s,v_t),(v_s,w_t)\}$
 all in the set $\mathcal{F}$. By the definition of the cartesian product of the weighted graphs [2], each of these edges is the cartesian product of the vertex $v_s$ in say $(G_1,b)$ with  the edge $\{v_t,w_t\}$, in say $(G_2,c)$ where $a(\{(v_s,v_t),(v_s,w_t)\}) =\mathfrak{d}_{v_s} c(\{v_t,w_t\}).$  Thus the graph $(G_2,c)$ is a graph of $q$ vertices obtained by projecting each of the $q(q-1)/2$ edges $\{(v_s,v_t),(v_s,w_t)\}$ to $\{v_t,w_t\}$, with corresponding weight assignments, and thus is a complete graph on $q$ vertices. As we vary $v_s$ through its $p$ possible values, one row corresponding to each value,  the definition of the cartesian product of weighted graphs generates the same $(G_2,c) $ possibly with different weights on edges. 
In exactly the same way, $q$ columns in figure 1 generate the complete graph on $p$ vertices $(G_1,b)$ by employing the cartesian product of weighted graphs. Noting that cartesian product of an isolated vertex in $(G_1,b)$ with an edge in $(G_2,c)$ or vice versa gives an isolated vertex in the product graph,  we have shown that the spanning subgraph of $(G,a)$ corresponding to the $\mathcal{F}$ set edges gets generated by $(G_s,b) \square (G_t,c)$ containing $d_s$ and $d_t$ vertices respectively.  Finally note that there are no loops in $(G_s,b) \square (G_t,c)$ because loops contribute to the cartesian product of weighted graphs only via the degrees $\mathfrak{d}_{v_s}$ and $\mathfrak{d}_{v_t}$. 
In  the above analysis, question can be raised regarding the weight functions of the factors. Thus more than one weight functions can generate the spanning subgraph corresponding to the $\mathcal{C}$ set while the $\mathcal{F}$ set edges on different rows in figure 1 may be generated by different weight functions on the factors in the cartesian product. These points will be addressed later in this proof.
We  denote by $G_1$ and $G'_1$  to be the graphs underlying  $(G_1,b)$ and $(G'_1,b')$ and similarly define $G_2$,  $G'_2$, $G$, $G'_1 \otimes G'_2$ and $G_1 \square G_2$. We first note that $|V(G_1)| =p=|V(G'_1)|$ and $|V(G_2)|=q=|V(G'_2)|$. In fact, we can identify the set of vertices for $ G_1$ and $G'_1$ and for $G_2$ and $G'_2$ respectively, that is $V(G_1)= V(G'_1)$ and $V(G_2)= V(G'_2)$. We now show that the identity map on $V(G_1)$ is an automorphism taking $G_1$ to $G'_1$. In other words, $G_1$ and $G'_1$ are identical.  Consider  a vertex $(v_s,v_t)$ in $G$ and let $N_G(v_s,v_t)$ denote its neighborhood in $G$. We denote by $N_{G_1 \square G_2}(v_s,v_t)$ and $N_{G'_1 \otimes G'_2}(v_s,v_t)$  the neighborhoods of $(v_s,v_t)$ in $G_1 \square G_2$ and $G'_1 \otimes G'_2$ respectively.
 In other words $N_{G'_1 \otimes G'_2}(v_s,v_t)$ contains neighbors of $(v_s,v_t)$ with edges in set $\mathcal{C}$  and  $N_{G_1 \square G_2}(v_s,v_t) $  contains the neighbors of $ (v_s,v_t)$ with edges in set   $\mathcal{F}$. Clearly $N_{G'_1 \otimes G'_2}(v_s,v_t)$ and   $N_{G_1 \square G_2}(v_s,v_t) $  partition  $N_G(v_s,v_t)$.
 Let $V_K(G_1)$ and $V_K(G_2)$ be the set of vertices on the clique in $G_1$ and $G_2$ respectively.  Consider [14]
 
  $$N_{G'_1 \otimes G'_2}(v_s,v_t)= V_K(G) - \{(v_s,v_t)\} - N_{G_1 \square G_2}(v_s,v_t) $$

$ =V_K(G) - \{v_s\} \times \{v_t\} -\{ \{v_s\} \times  N_{G_2}(v_t) \cup N_{G_1} (v_s) \times \{v_t\}\} $

 $ =V_K(G_1) \times V_K(G_2) - \{v_s\} \times \{v_t\} -\{ \{v_s\} \times  N_{G_2}(v_t) \cup N_{G_1} (v_s) \times \{v_t\}\} $
\\

 $ =\{N_{G_1} (v_s) \cup \{v_s\} \} \times \{N_{G_2}(v_t) \cup \{v_t\}\} - \{v_s\} \times \{v_t\} -\{ \{v_s\} \times  N_{G_2}(v_t) \cup N_{G_1} (v_s) \times \{v_t\}\} $
\\

$ =N_{G_1} (v_s) \times N_{G_2}(v_t)  \cup N_{G_1} (v_s) \times  \{v_t\}  \cup \{v_s\}  \times N_{G_2}(v_t) \cup\{v_s \} \times \{v_t\} - \{v_s\} \times \{v_t\} $
\\

    $  -\{ \{v_s\} \times  N_{G_2}(v_t) \cup N_{G_1} (v_s) \times \{v_t\}\} $
\\

$ =N_{G_1} (v_s) \times N_{G_2}(v_t) =N_{G_1 \otimes G_2}(v_s,v_t). $
\\

Therefore [14],  $ N_{G'_1} (v_s) \times N_{G'_2}(v_t) = N_{G_1} (v_s) \times N_{G_2}(v_t)$
\\

or  $N_{G'_1} (v_s) =N_{G_1} (v_s) $  and $N_{G'_2}(v_t) = N_{G_2}(v_t).$

Therefore,  $G_1$ and $G'_1$ and $G_2$ and $G'_2$  are identical.  In particular $G_1$ and $G_2$ are cliques. $G_s=G'_s$ and $G_t=G'_t$ are cliques plus isolated vertices.  Thus we get  
$$ (G,a) =\cL(G_s,b') \otimes \cL(G_t,c') \dotplus(G_s,b)\square (G_t,c) \dotplus (G'',a) \eqno{(3)}$$
where $(G'',a)$ is the graph obtained from $(G,a)$ by removing all edges and keeping loops.
\\

The only remaining gap is to show that a consistent assignment of weights to the factors $G_s$ and $G_t$ is possible so as to express $(G,a)$ as modified tensor product. To get the required weight assignments we use the requirement that both the factors in the modified tensor product must correspond to pure states.  Indeed, we know that both $G_s$ and $G_t$ have the form of clique plus isolated vertices as required for them to represent pure states. We know that the graph $(G,a)$ corresponds to pure state. Therefore, its weight function $a$ must satisfy [2], {\it assuming $(G,a)$ to be a real weighted graph},

$$\sum_{(v_s,v_t) \in V(G,a)} \mathfrak{d}^2_{(v_s,v_t)} + 2\sum_{e \in E(G,a))} a^2(e) = \mathfrak{d}^2_{(G,a)}.  \eqno{(4)}$$

Splitting $E(G,a)$ into   $\mathcal{C}$ and $\mathcal{F}$ sets and using the definitions of the tensor and cartesian products of weighted graphs, we get, (note that a paired label is for a vertex in $V(G,a)$ and single labels with suffix $s$ and $t$ are vertices in $V(G_s,b)$ and $V(G_t,c)$ respectively),
$$
\sum_{(v_s,v_t) } \mathfrak{d}^2_{(v_s,v_t)} + 2 \sum_{(v_s,v_t) } 
 \sum_{\begin{subarray}{I}          
     \hskip  .5cm       (w_s,w_t)\\                
     \hskip .2 cm { w_s \ne v_s , w_t \ne v_t}        
   \end{subarray}} 							
 (a^2(\{(v_s,v_t),(w_s,w_t)\}) +a^2(\{(v_s,w_t),(w_s,v_t)\}))	
$$
  $$ 
+ 2 \sum_{(v_s,v_t)}  
\sum_{\begin{subarray}{|}
  \hskip .2 cm  {(v_s,w_t)} \\
   \hskip .2 cm {  w_t \ne v_t} 
\end{subarray}} 
a^2(\{(v_s,v_t),(v_s,w_t)\})
+2 \sum_{(v_s,v_t) } 
 \sum_{\begin{subarray}{|}
     \hskip .2 cm (w_s,v_t)\\
    \hskip .2 cm { w_s \ne v_s } 
\end{subarray}} 
a^2(\{(v_s,v_t),(w_s,v_t)\}) = \mathfrak{d}^2_{(G,a)}.
$$

 Since $E(G,a)$ is closed under $T_s$  we can write

$$
\sum_{(v_s,v_t) } \mathfrak{d}^2_{(v_s,v_t)} + 4 \sum_{(v_s,v_t) } 
\sum_{\begin{subarray}{I} 
  \hskip  .5cm   (w_s,w_t) \\
   \hskip  .2cm   { w_s \ne v_s , w_t \ne v_t}
\end{subarray}} 
 a^2(\{(v_s,v_t),(w_s,w_t)\})
$$ 
 $$ 
+ 2 \sum_{(v_s,v_t) }  
\sum_{\begin{subarray}{I} 
 \hskip  .1cm   (v_s,w_t) \\
 \hskip  .1cm   {  w_t \ne v_t} 
\end{subarray}} 
a^2(\{(v_s,v_t),(v_s,w_t)\}) 
+2 \sum_{(v_s,v_t) }  
\sum_{\begin{subarray}{I} 
 \hskip  .1cm   (w_s,v_t)\\
 \hskip  .1cm   { w_s \ne v_s }
\end{subarray}} 
 a^2(\{(v_s,v_t),(w_s,v_t)\}) = \mathfrak{d}^2_{(G,a)}. \eqno{(5)}$$

 Using splitting of the weight functions in $\mathcal{C}$ set and $\mathcal{F}$ set we get

$$
\sum_{(v_s,v_t) } \mathfrak{d}^2_{(v_s,v_t)} + 4 \sum_{v_s } 
\sum_{\begin{subarray}{I} 
\hskip  .4 cm {w_s }\\
\hskip  .1cm { w_s \ne v_s } 
\end{subarray}} 
b'^2(\{v_s,w_s\}) \sum_{v_t }  
\sum_{\begin{subarray}{I}
\hskip  .4cm {w_t }\\
\hskip  .1cm { w_t \ne v_t }
\end{subarray}} 
c'^2(\{v_t,w_t\}) 
$$

 $$
 + 2 \sum_{v_s }  \mathfrak{d}^2_{v_s} \sum_{v_t } 
\sum_{\begin{subarray}{I}
\hskip  .4cm{w_t }\\  
\hskip  .1cm{w_t \ne v_t} 
\end{subarray}} 
 c^2(\{v_t,w_t\} 
+2 \sum_{v_t } \mathfrak{d}^2_{v_t}  \sum_{v_s }   
 \sum_{\begin{subarray}{I}
\hskip  .4cm {w_s }\\  
\hskip  .1cm {w_s \ne v_s} 
\end{subarray}} 
 b^2(\{v_s,w_s\}) =\mathfrak{d}^2_{(G,a)}. \eqno{(6)}$$   

Since the graphs $(G_s,b),(G_t,c) ,(G_s,b')$, and $(G_t,c')$  correspond to pure states,  $b' , c', b ,c $ must satisfy,

$$
\sum_{v_s} \mathfrak{d}^2_{v_s} + 2 \sum_{v_s }  
\sum_{\begin{subarray}{I}
\hskip  .4cm {w_s}\\
\hskip  .1cm { w_s \ne v_s } 
\end{subarray}} 
b^2(\{v_s,w_s\})= \mathfrak{d}^2_{(G_s,b)}\eqno{(7)}$$

$$
\sum_{v_s } \mathfrak{d}^2_{v_s} + 2 \sum_{v_s }  
\sum_{\begin{subarray}{I}
\hskip  .4cm {w_s }\\
\hskip  .1cm { w_s \ne v_s }
\end{subarray}} 
 b'^2(\{v_s,w_s\})= \mathfrak{d}^2_{(G_s,b')}\eqno{(8)}$$

$$
\sum_{v_t } \mathfrak{d}^2_{v_t} + 2 \sum_{v_t } 
 \sum_{\begin{subarray}{I}
\hskip  .4cm {w_t }\\
\hskip  .1cm{ w_t \ne v_t } 
\end{subarray}} 
c^2(\{v_t,w_t\})= \mathfrak{d}^2_{(G_t,c)}\eqno{(9)}$$

$$
\sum_{v_t } \mathfrak{d}^2_{v_t} + 2 \sum_{v_t} 
 \sum_{\begin{subarray}{I}
\hskip  .4cm {w_t}\\
\hskip  .1cm { w_t \ne v_t } 
\end{subarray}} 
c'^2(\{v_t,w_t\})= \mathfrak{d}^2_{(G_t,c')}.\eqno{(10)}$$

We see that Eqs. (7), (8), (9), and (10) are consistent with Eq. (6) provided 
\\

$(i) \; \mathfrak{d}_{(v_s,v_t)}  = \mathfrak{d}_{v_s} \mathfrak{d}_{v_t}$ for all $(v_s,v_t) \in V(G,a)$ and consequently $\mathfrak{d}_{(G,a)} = \mathfrak{d}_{(G_s,b)} \mathfrak{d}_{(G_t,c)}$
\\

$ (ii)  \; b^2(\{v_s,w_s\}) = b'^2(\{v_s,w_s\}) ;  \; c^2(\{v_t,w_t\}) = c'^2(\{v_t,w_t\})$.
\\

We first fix a vertex $(v_s,v_t) \in V(G,a)$ and obtain its degree $ \mathfrak{d}_{(v_s,v_t)}$.  Summing the edges with the same $s$ part we get 

$$
\sum_{\begin{subarray}{I}
\hskip  .1cm (v_s,w_t)\\
\hskip  .1cm{ w_t \ne v_t}
\end{subarray}} 
 a(\{(v_s,v_t),(v_s,w_t)\}) = \mathfrak{d}_{v_s} 
\sum_{\begin{subarray}{I}
\hskip  .4cm {w_t} \\
\hskip  .1cm { w_t \ne v_t } 
\end{subarray}} 
c(\{v_t,w_t\}). \eqno{(11)}
$$

Adding edges with the same $t$ part we have 

$$
\sum_{\begin{subarray}{I}
\hskip  .1cm{(v_s,w_t)}\\
\hskip  .1cm{  w_s \ne v_s}
\end{subarray}} 
 a(\{(v_s,v_t),(w_s,v_t)\}) = \mathfrak{d}_{v_t} 
\sum_{\begin{subarray}{I}
\hskip  .4cm{w_s }\\
\hskip  .1cm{ w_s \ne v_s } 
\end{subarray}} 
b(\{v_s,w_s\}). \eqno{(12)}$$

Adding over edges in the  $\mathcal{C}$ set we get 

$$
\sum_{\begin{subarray}{I}
\hskip  .1cm (v_s,w_t) \\
\hskip  .1cm {  w_t \ne v_t}
\end{subarray}} 
 a(\{(v_s,v_t),(w_s,w_t)\}) =
\sum_{\begin{subarray}{I}
\hskip  .4cm {w_s}\\
\hskip  .1cm{ w_s \ne v_s } 
\end{subarray}} 
b'(\{v_s,w_s\})  
\sum_{\begin{subarray}{I}
\hskip  .4cm {w_t}\\
\hskip  .1cm { w_t \ne v_t } 
\end{subarray}} 
c'(\{v_t,w_t\}). \eqno{(13)}$$

Adding these three terms and the weight of the loop on $(v_s,v_t)$ we get $\mathfrak{d}_{(v_s,v_t)}$. The requirement that $\mathfrak{d}_{(v_s,v_t)} = \mathfrak{d}_{v_s} \mathfrak{d}_{v_t}$ is  satisfied provided
\\

$(iii) \; b'(\{v_s,w_s\}) = b(\{v_s,w_s\})$ and $c'(\{v_t,w_t\}) = -c(\{v_t,w_t\})$  which leads to 
\\

$(iv) \; \mathfrak{d}_{(v_s,v_t)} - a(\{(v_s,v_t),(v_s,v_t)\}) = \mathfrak{d}_{v_s} \mathfrak{d}_{v_t} - b(\{v_s,v_s\}) c(\{v_t,v_t\}).$
\\

 This is satisfied provided $ a(\{(v_s,v_t),(v_s,v_t)\}) = b(\{v_s,v_s\}) \cdot c(\{v_t,v_t\})$. This requirement is consistent with $ (G'',a) = \Om (G_s,b) \otimes \Om(G_t,c).$

We can write, therefore, $$ (G, a)   =  \cL(G_s, b) \otimes \cL  (G_t,- c) \dotplus (G_s, b) \square (G_t,c)\\
\dotplus  \Om(G_s, b) \otimes \Om(G_t,c)$$  $$ = (G_s,b) \boxdot (G_t,c)$$

Now, let us deal with the case where $(G,a)$\textit{ is a graph with complex weights} [2]. In this case Eq. (4) is replaced by  

$$\sum_{(v_s,v_t) \in V(G,a)} \mathfrak{d}^2_{(v_s,v_t)} + 2\sum_{e \in E(G,a))} |a(e)|^2 = \mathfrak{d}^2_{(G,a)}  \eqno{(4b)}$$

and Eqs. (5, 6, 7, 8, 9, 10)  become 

$$
\sum_{(v_s,v_t) } \mathfrak{d}^2_{(v_s,v_t)} + 4 \sum_{(v_s,v_t) } 
\sum_{\begin{subarray}{I} 
  \hskip  .5cm   (w_s,w_t) \\
   \hskip  .2cm   { w_s \ne v_s , w_t \ne v_t}
\end{subarray}} 
| a(\{(v_s,v_t),(w_s,w_t)\})|^2
$$ 
 $$ 
+ 2 \sum_{(v_s,v_t) }  
\sum_{\begin{subarray}{I} 
 \hskip  .1cm   (v_s,w_t) \\
 \hskip  .1cm   {  w_t \ne v_t} 
\end{subarray}} 
|a(\{(v_s,v_t),(v_s,w_t)\})|^2
+2 \sum_{(v_s,v_t) }  
\sum_{\begin{subarray}{I} 
 \hskip  .1cm   (w_s,v_t)\\
 \hskip  .1cm   { w_s \ne v_s }
\end{subarray}} 
| a(\{(v_s,v_t),(w_s,v_t)\})|^2 = \mathfrak{d}^2_{(G,a)} \eqno{(5b)}$$
\\
 
$$
\sum_{(v_s,v_t) } \mathfrak{d}^2_{(v_s,v_t)} + 4 \sum_{v_s } 
\sum_{\begin{subarray}{I} 
\hskip  .4 cm {w_s }\\
\hskip  .1cm { w_s \ne v_s } 
\end{subarray}} 
|b'(\{v_s,w_s\})|^2 \sum_{v_t }  
\sum_{\begin{subarray}{I}
\hskip  .4cm {w_t }\\
\hskip  .1cm { w_t \ne v_t }
\end{subarray}} 
|c'(\{v_t,w_t\})|^2 
$$
 $$
 + 2 \sum_{v_s }  \mathfrak{d}^2_{v_s} \sum_{v_t } 
\sum_{\begin{subarray}{I}
\hskip  .4cm{w_t }\\  
\hskip  .1cm{w_t \ne v_t} 
\end{subarray}} 
 |c(\{v_t,w_t\} )|^2
+2 \sum_{v_t } \mathfrak{d}^2_{v_t}  \sum_{v_s }   
 \sum_{\begin{subarray}{I}
\hskip  .4cm {w_s }\\  
\hskip  .1cm {w_s \ne v_s} 
\end{subarray}} 
 |b(\{v_s,w_s\})|^2 =\mathfrak{d}^2_{(G,a)} \eqno{(6b)}$$   
\\

$$
\sum_{v_s} \mathfrak{d}^2_{v_s} + 2 \sum_{v_s }  
\sum_{\begin{subarray}{I}
\hskip  .4cm {w_s}\\
\hskip  .1cm { w_s \ne v_s } 
\end{subarray}} 
|b(\{v_s,w_s\})|^2 = \mathfrak{d}^2_{(G_s,b)}\eqno{(7b)}$$
\\

$$
\sum_{v_s } \mathfrak{d}^2_{v_s} + 2 \sum_{v_s }  
\sum_{\begin{subarray}{I}
\hskip  .4cm {w_s }\\
\hskip  .1cm { w_s \ne v_s }
\end{subarray}} 
| b'(\{v_s,w_s\})|^2 = \mathfrak{d}^2_{(G_s,b')}\eqno{(8b)}$$
\\

$$
\sum_{v_t } \mathfrak{d}^2_{v_t} + 2 \sum_{v_t } 
 \sum_{\begin{subarray}{I}
\hskip  .4cm {w_t }\\
\hskip  .1cm{ w_t \ne v_t } 
\end{subarray}} 
|c(\{v_t,w_t\})|^2 = \mathfrak{d}^2_{(G_t,c)}\eqno{(9b)}$$
\\

$$
\sum_{v_t } \mathfrak{d}^2_{v_t} + 2 \sum_{v_t} 
 \sum_{\begin{subarray}{I}
\hskip  .4cm {w_t}\\
\hskip  .1cm { w_t \ne v_t } 
\end{subarray}} 
|c'(\{v_t,w_t\})|^2= \mathfrak{d}^2_{(G_t,c')}.\eqno{(10b)}$$

We see that Eqs. (7b), (8b), (9b), and (10b) are consistent with Eq. (6b) provided 
\\

$(v)\;  \mathfrak{d}_{(v_s,v_t)}  = \mathfrak{d}_{v_s} \mathfrak{d}_{v_t}$ for all $(v_s,v_t) \in V(G,a)$ and consequently $\mathfrak{d}_{(G,a)} = \mathfrak{d}_{(G_s,b)} \mathfrak{d}_{(G_t,c)}$
\\

$ (vi)\; |b(\{v_s,w_s\})|^2\; =\; |b'(\{v_s,w_s\})|^2;  \; |c(\{v_t,w_t\})|^2 \; =\; |c'(\{v_t,w_t\})|^2$.
\\

 Eqs. (11, 12, 13)  become

$$\sum_{\begin{subarray}{I}
(v_s,w_t) \\
{ w_t \ne v_t} 
\end{subarray}} 
|a(\{(v_s,v_t),(v_s,w_t)\})| \; =\; \mathfrak{d}_{v_s} 
\sum_{\begin{subarray}{I}
\hskip  .4cm {w_t} \\
{ w_t \ne v_t }
\end{subarray}} 
 |c(\{v_t,w_t\})|  \eqno{(11b)}$$
\\

$$\sum_{\begin{subarray}{I}
(v_s,w_t) \\
{  w_s \ne v_s}
\end{subarray}} 
 |a(\{(v_s,v_t),(w_s,v_t)\})| \; =\; \mathfrak{d}_{v_t}
 \sum_{\begin{subarray}{I}
\hskip  .4cm {w_s }\\
{ w_s \ne v_s }
\end{subarray}} 
 |b(\{v_s,w_s\})|  \eqno{(12b)}$$
\\

$$\sum_{\begin{subarray}{I}
(v_s,w_t)\\
{ w_t \ne v_t} 
\end{subarray}} 
|a(\{(v_s,v_t),(w_s,w_t)\})| 
=\sum_{\begin{subarray}{I}
\hskip  .4cm {w_s }\\
{ w_s \ne v_s } 
\end{subarray}} 
|b'(\{v_s,w_s\})|
  \sum_{\begin{subarray}{I}
\hskip  .4cm {w_t }\\
{ w_t \ne v_t }
\end{subarray}} 
 |c'(\{v_t,w_t\})|.  \eqno{(13b)}$$

Adding these three terms and the weight of the loop on $(v_s,v_t)$ we get $\mathfrak{d}_{(v_s,v_t)}$. The requirement that $\mathfrak{d}_{(v_s,v_t)} = \mathfrak{d}_{v_s} \mathfrak{d}_{v_t}$ is  satisfied provided 
\\

$(vii) \; |b'(\{v_s,w_s\})| \;=\; |b(\{v_s,w_s\})| $ and  $ |c'(\{v_t,w_t\})| \;=\; |c(\{v_t,w_t\})|$
\\

$$(viii) \; a(\{(v_s,v_t),(v_s,v_t)\})\;=\; 
b(\{v_s,v_s\}) c(\{v_t,v_t\})\;-\; 2
 \sum_{\begin{subarray}{I}
\hskip  .4cm {w_s}\\
{ w_s \ne v_s }
\end{subarray}} 
 |b(\{v_s,w_s\})|  
\sum_{\begin{subarray}{I}
\hskip  .4cm {w_t}\\
{ w_t \ne v_t }
\end{subarray}} 
 |c(\{v_t,w_t\})|.$$

The requirement $(viii)$ is consistent with 
$$(G^{''},a)\;=\;  \Om(G_s, b) \otimes \Om(G_t,c) \sqcup 2 \cN \cL (G_s,b) \otimes \cN \cL \eta (G_t,c). \eqno{(14)}$$

We now choose phases of the weight functions $b$ and  $c$.  Consider the edges $e=\{(v_s,v_t),(w_s,w_t)\}$ and $T_s e \;=\;\{(w_s,v_t),(v_s,w_t)\}$  in $E(G,a)$. We know that $|a(e)| \;=\; |a(T_s e)|$ . Let $ e^{i \th_1} $ and $ e^{i \th_2}$ be the phases of $a_1=a(e)$ and $a_2= a(T_s e)$ respectively. If we require that these two edges in $E(G,a)$ be produced by the tensor product of the edge $\{v_s,w_s\} $ in $(G_s,b)$ with the edge $\{v_t,w_t\} $ in $(G_t,c)$  (see figure 2) then the phases of weights $b$ and $c$ on the corresponding edges must be $\phi_1 \;=\; (\th_1 + \th_2 )/2$ and $ \phi_2 \;=\; (\th_1 - \th_2 )/2$ respectively.

\begin{figure}[!ht]
\begin{center}
\includegraphics[width=16cm,height=4cm]{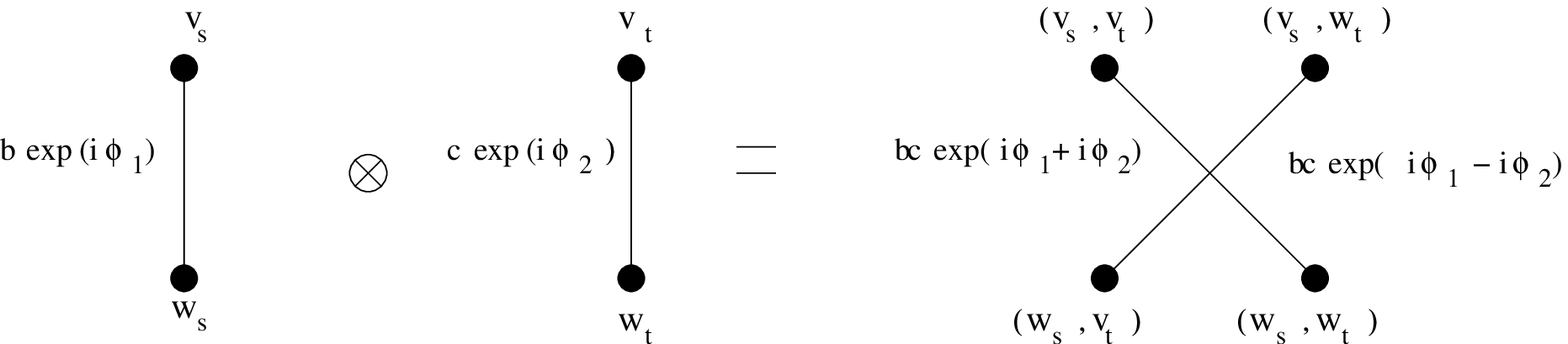}

figure 2 
\end{center}
\end{figure}

This completely fixes the weight functions $b$ and $c$ on $(G_s,b)$ and $(G_t,c)$ respectively. We now have, for every edge $e$ in $E(G,a)$ the corresponding edges $e_1,e_2$ in $(G_s,b)$ and $(G_t,c)$ respectively such that $ e=e_1 \otimes e_2$ and $a(e)=b(e_1) c(e_2)$. Thus we have, finally from Eq. (3), $(vii)$ and (14)

$$ (G,a)\; =\; \cL(G_s,b) \otimes \cL (G_t, c) \dotplus \cL(G_s,b) \otimes \cN(G_t,c) \dotplus \cN(G_s,b) \otimes \cL(G_t,c)$$
  $$ \dotplus \{ \Om(G_s, b) \otimes \Om(G_t,c) \sqcup 2 \cN \cL (G_s,b) \otimes \cN \cL \eta (G_t,c)\}  \eqno{(2)}$$ 
$$\;=\;(G_s,b) \boxdot (G_t,c).$$
\hfill $\blacksquare$ 
\\

 \textbf{ Lemma 2.3} :\textit{ Let $ \si , \si_s $ and $ \si_t$ be density matrices for pure states. Then $ \si = \si_s \otimes \si_t$ if and only if $(G,a) = (G_s,b) \boxdot (G_t,c)$ where $(G,a),\; (G_s,b) $ and  $(G_t,c)$ are the graphs for  $ \si,\; \si_s $ and $ \si_t$ respectively.}\\

\textit{Proof :}  This Lemma is identical with theorems  4.5, 6.10 in [2].  \hfill $\blacksquare$ \\

\textbf{Theorem 2.1} : \textit{Let $(G,a)$ be the graph of a $m$-partite pure state $\si$ in the Hilbert space $\mathcal{H}^{d_1}\otimes \mathcal{H}^{d_2} \otimes \cdots \otimes \mathcal{H}^{d_m}$.  Let $\mathcal{H}^{(s)}=\mathcal{H}^{d_{i_1}} \otimes \mathcal{H}^{d_{i_2}} \otimes \mathcal{H}^{d_{i_3}}\otimes \cdots \otimes \mathcal{H}^{d_{i_s}},\; \{ i_1 ,i_2,\cdots ,i_s\} = s$  correspond to an  $s$ set,  $\mathcal{H}^{(t)}=\mathcal{H}^{d_{j_1}} \otimes \mathcal{H}^{d_{j_2}} \otimes \mathcal{H}^{d_{j_3}}\otimes \cdots \otimes \mathcal{H}^{d_{j_t}},\;  \{j_1,j_2, \cdots ,j_t  \} =t$, correspond to the $t$ set which is the complement of $s$ set in $\{1,2, \cdots,m\}$. Then $\si = \si_s \otimes \si_t$,  where 
 $\si_s $ and $ \si_t$ are pure states in $\mathcal{H}^{(s)}$ and $\mathcal{H}^{(t)}$  with graphs $(G_s,b)$ and $(G_t,c)$  respectively, if and only if $\Delta (G,a) =\Delta (G^{T_s},a')$.}\\

\textit{Proof :} Using Lemmas 2.1, 2.2 and 2.3 we have   

$\si = \si_s \otimes \si_t  \Longleftrightarrow  (G,a) = (G_s,b) \boxdot (G_t,c)  \Longleftrightarrow  E(G,a)  $ is closed under $ T_s \Longleftrightarrow  \Delta (G,a) =\Delta (G^{T_s},a')$. A state $\si$ is entangled if  $ \Delta (G,a) \ne \Delta (G^{T_s},a')$  in every partition $s$ and $t$ of $\{1,2, \cdots,m\}$. \hfill $\blacksquare$\\

\textbf{III. ALGORITHM}
\\

While proving the if part of Lemma 2.2, we have shown that the number of vertices in the cliques of the factors $G_1$ and $G_2$ defined there ($p$ and $q$ respectively) are the factors of the number of vertices on the clique in $(G,a)$, that is, $n=pq$. This means that the $m$-partite pure state $|\psi \ran$ corresponding to $(G,a)$ has two factors   $|\psi_1 \ran$  and  $|\psi_2 \ran$, corresponding to $G_1$ and $G_2$ respectively, such that  $|\psi_1 \ran$ lives in a $p$-dimensional subspace of $\mathcal{H}^{d_{i_1}} \otimes \mathcal{H}^{d_{i_2}} \otimes \mathcal{H}^{d_{i_3}}\otimes \cdots \otimes \mathcal{H}^{d_{i_s}}$ and $|\psi_2 \ran$ lives in a $q$-dimensional subspace of $\mathcal{H}^{d_{j_1}} \otimes \mathcal{H}^{d_{j_2}} \otimes \mathcal{H}^{d_{j_3}}\otimes \cdots \otimes \mathcal{H}^{d_{j_t}}$. If  the weighted versions of  $G_1$ and $G_2$, namely $(G_s,b)$ and $(G_t,c)$, can be further factorized, the dimensions of the corresponding subspaces will be the factors of $p$ and $q$ respectively. This procedure can be iterated at most until the dimensions of the subspaces for the factors of $|\psi \ran$  are the prime factors of $n$. Therefore, the dimension of the subspaces containing the factors of $|\psi \ran$ are the prime factors of $n$ or the products of such factors. This fact can be used to get a polynomial algorithm to find the full separability of $|\psi \ran$ in the following way. By full separability we mean expressing $|\psi \ran$ as a product state whose further factorization is impossible. Denote by $p_1 \geq p_2 \geq \cdots \geq p_k$ the prime factors of $n$. Let $d_1 \ge d_2 \ge d_3 \ge \cdots \ge d_m $ be the dimensions of the Hilbert spaces of $m$ parts arranged in a non-increasing order. Let $s_1 > 0$ be the least integer satisfying $d_1 d_2 \cdots d_{s_1} \ge p_1$. We implement our algorithm (Theorem 2.1) on partitions $(s,t)$ with $s_1\leq s \leq m-1$. The total number of times the algorithm has to run, in the worst case, is 
\begin{displaymath}
\binom{m}{s_1}\;+\;\binom{m}{s_1+1}\;+\; \cdots+\;\binom{m}{m-1}
\end{displaymath}

 which is a polynomial of degree $s_1$ in $m$. Thus we have a polynomial algorithm to check separability of a m-partite system. Suppose we get the separability as $\mathcal{H}^{(s)}\otimes \mathcal{H}^{(t)}$. Then the factor in $\mathcal{H}^{(s)}$ cannot be further factorized as it corresponds to the largest prime factor of $n$ and $\mathcal{H}^{(t)}$ contains factors corresponding to $ p_2 \geq p_3 \cdots \geq p_k$. We repeat the above algorithm on $\mathcal{H}^{(t)}$ with $p_2$ as the largest prime factor. Its worst case complexity is given by a polynomial of degree $(m-s)^{s_2}$ where $s_2$ is defined like $s_1$ above. We carry out these iterations until full separability is obtained. Thus if we do not get any factorization in the first iteration, corresponding to the largest prime factor $p_1$, then the state is fully entangled, like GHZ or W state. Unless the factorization carries up to $m$ factors, the factors of the state contain one or more entangled states involving less than $m$ parts. The total algorithm is polynomial in $m$. Note that, if $n$ is prime, then all that is necessary is to look for some $v_i$ common to the $m$ tuples for {\it all} vertices on the clique. If say $v_i$ is common, then $$| \psi \ran\;=\;|\phi\ran\otimes |v_i\ran$$ with $|\phi\ran \in \mathcal{H}^{d_1}\otimes \cdots \otimes  \mathcal{H}^{d_{i-1}}\otimes \mathcal{H}^{d_{i+1}}\otimes \cdots \otimes \mathcal{H}^{d_m}$ and $|v_i \ran \in \mathcal{H}^{d_i}$. Otherwise the given state $| \psi \ran$ is entangled.\\

\textbf{IV. A COUNTER EXAMPLE }\\

 We  note that Theorem 2.1 may not apply to mixed states  as the following example shows. Consider the bipartite separable state $\si\;=\; 1/2 |y,- \ran |y,- \ran \lan y,-| \lan y,-| + 1/2 |x,+ \ran |x,+ \ran \lan x,+| \lan x,+|$, where $|y,- \ran = \fr{1}{ \sqrt2} (|0 \ran - i |1 \ran)$ and $   |x,+ \ran = \fr{1}{ \sqrt2} (|0 \ran + |1 \ran)$. The corresponding density matrix in standard  basis is $$\si = \fr{1}{4} \left[ \ba{rrrr} 2 & 1+i & 1+i & 0 \\1-i & 2 & 2 & 1+i \\1-i & 2 & 2 & 1+i \\ 0 &1-i &1-i & 2 \ea \right] $$ and  the corresponding graph is shown in figure 3. We see that the $\mathcal{C}$ set contains only one edge  for all possible partitions and hence cannot be closed under any $T_s$. We show, in section (V), that the degree conjecture applies to states with real weighted graphs without loops. Therefore, the above example shows that the degree conjecture does not apply to all mixed states.
 \\

\begin{figure}[!ht]
\begin{center}
\includegraphics[width=6cm,height=5cm]{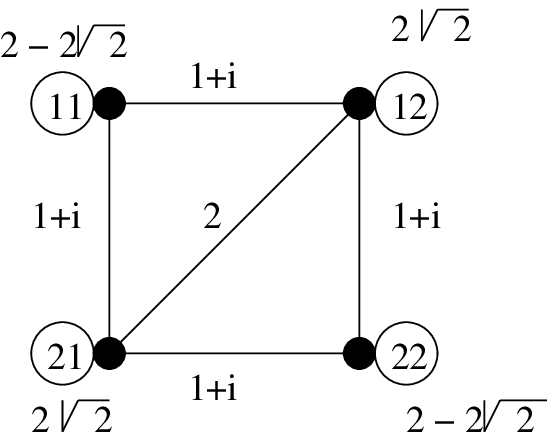}

figure 3 
\end{center}
\end{figure}

\textbf{V. PROOF OF DEGREE CRITERION FOR A CLASS OF MULTIPARTITE MIXED STATES}\\

In this section we prove the degree criterion for the class of states whose graphs have no loops and have real weights.\\

\textbf{Theorem 5.1 :} Let $\si(G,a)$ be a density matrix acting on $\mathcal{H}^{d_1} \otimes \mathcal{H}^{d_2} \otimes \cdots \otimes \mathcal{H}^{d_m}$, with a real weighted graph $(G,a)$, on $d=d_1 d_2 \dots d_m$ vertices,  having no loops. If $\si(G,a)$ is separable in $s,t$ cut where $s,t$ is a partition of $\{1,2,\dots,m\}$, so that $\si(G,a)=\sum_i p_i \si_i^{(s)}\otimes \si_i^{(t)}$, where  $\sigma_i^{(s)}$ and $ \sigma_i^{(t)}$ are density matrices acting on $\mathcal{H}^{(s)}$ and $\mathcal{H}^{(t)}$ with graphs $(G_i^{(s)},b)$ and $(G_i^{(t)},c)$ respectively, then $\Delta(G,a)=\Delta(G^{T_s},a')$.\\

{\bf Proof :} Let $Q(G,a)$ be the Laplacian of a graph $(G,a)$ with  real  weights without  loops, on $d$ vertices. For a graph without loops $Q(G,a)=L(G,a)$ [2]. Let $D$ be any $d\times d$ real diagonal matrix in the standard orthonormal basis $\{|v_i\ran\}; \; i=1,2,\dots,d$, such that $D\neq 0$ and $Tr(D)=0$, where $Tr$ is the trace of $D$. This means that there is at least one negative entry in the diagonal of $D$. Denote this element by $D_{ii}=b_i < 0$. Let $|\psi_0 \ran= \sum_j|v_j\ran$ and $|\phi\ran=\sum_j \chi_j|v_j \ran$  where, 
\begin{displaymath}
\chi_j = 
\left\{ \begin{array}{ll}
 0 & \textrm{if $j\neq i$}\\
k  & \textrm{if $j=i$},\; \mbox{$k$ real.}
\end{array} \right.
\end{displaymath}
Let $|\chi \ran=|\psi_0\ran+|\phi\ran=\sum_{j=1}^d(1+\chi_j)|v_j\ran$. Then 
\benrr
\lan\chi|L(G,a)+D|\chi\ran& = & \lan\psi_0|L(G,a)|\psi_0\ran+
\lan\psi_0|L(G,a)|\phi\ran+\lan\phi|L(G,a)|\psi_0\ran+\\
& &              \lan\phi|L(G,a)|\phi\ran+\lan\psi_0|D|\psi_0\ran+\lan\psi_0|D|\phi\ran+\lan\phi|D|\psi_0\ran+\lan\phi|D|\phi\ran ~~~~(15)
\eenrr

 Since $L(G, a)$ is positive semidefinite, we must have, for the quadratic form associated with $L(G,a)$ of a graph without loops [15,2], 
$$ x^T L(G, a)x = \sum_{\{j,l\}\in E(G,a)} a_{jl} (x_j - x_l)^2 \ge 0,  \eqno{(16)}$$
for every $x \in I\!\!R^d.$ 
Since $|\psi_0\ran$ is (unnormalized) vector having all components equal to unity, from Eq.(16) it follows that $\lan\psi_0|L(G,a)|\psi_0\ran=0$. Also $\lan\psi_0|D|\psi_0\ran=Tr(D)=0$. 
  We have
 $$\lan\phi|L(G,a)|\phi\ran=k^2(L(G,a))_{ii}=k^2\mathfrak{d}_i,$$
 where $\mathfrak{d}_i$ denotes the degree of $ith$ vertex. For a real weighted graph without loops the sum of the elements in any row of its Laplacian is zero. This leads to
 $$\lan\psi_0|L(G,a)|\phi\ran=\lan\phi|L(G,a)^T|\psi_0\ran = \lan\phi|L(G,a)|\psi_0\ran = 0.$$
 
 Finally, the remaining terms in Eq. (15) are given by 
 $$\lan\phi|D|\phi\ran=b_ik^2$$
 $$\lan\psi_0|D|\phi\ran = b_ik=\lan\phi|D|\psi_0\ran.$$
 Thus
 $$\lan\chi|L(G,a)+D|\chi\ran = k^2(b_i+\mathfrak{d}_i)+ 2kb_i$$
  So we can then always choose a positive $k$, such that
 $$\lan\chi|L(G,a)+D|\chi\ran < 0.$$
 Then it follows that $$L(G,a)+D \ngeq 0.\eqno{(17)}$$

  This expression is identical with that obtained in [1,2].
  For any graph $(G,a)$ on $d=d_sd_t$ vertices $$v_1=(u_1,w_1), v_2=(u_1,w_2), \dots, v_{d}=(u_{d_s},w_{d_t}),$$ consider  the degree condition $\Delta(G,a) = \Delta(G^{T_s},a^{\prime}).$  Since $M^{T_s}(G,a)=M(G^{T_s},a^{\prime})$ where $M$ is the adjacency matrix, we get,
  $$(L(G,a))^{T_s}=(\Delta(G,a) - \Delta(G^{T_s},a^{\prime}))+ L(G^{T_s},a^{\prime}).\eqno{(18)}$$
  Let $$D=\Delta(G,a) - \Delta(G^{T_s},a^{\prime}).$$
  
  Then $D$ is an $d\times d$ real diagonal matrix with respect to the orthonormal basis $$|v_i\ran = |u_1\ran\otimes |w_1\ran, \dots, |v_{d}\ran = |u_{d_s}\ran\otimes |w_{d_t}\ran.$$
  As the degree sum of a graph is invariant under partial transpose, 
  $$Tr(D)=Tr(\Delta(G,a))-Tr(\Delta(G^{T_s},a^{\prime}))= 0.$$
  We have two possible cases : $D\neq 0$ or $D = 0$. If $D\neq 0$, that is, the degree condition is not satisfied $(i.e.  \Delta(G,a)\neq \Delta(G^{T_s},a^{\prime}))$, we can write, via Eq.(17), 
   $L(G^{T_s},a^{\prime})+D\ngeq 0$ because, $(G^{T_s},a^{\prime})$ is a real weighted graph without loops and $D\ne 0$ with $Tr(D) = 0.$ Using Eq.(18), this means  $(L(G,a))^{T_s}\ngeq 0.$ As  $\si(G,a)=\fr{1}{\mathfrak{d}_{(G,a)}}L(G,a)$ [2], $\si(G,a)$ is entangled [5].
   
   \hfill $\blacksquare$\\

 In order to test the separability of the $m$-partite state $\si(G,a)$, we have to apply the degree criterion to all the bipartite cuts $(s,t)$ of $m$-partite system. This procedure may not detect the full separability of  $m$-partite state [16]. However, all the known separability criteria for multipartite states test separability only in bipartite cuts, and hence are not enough to guarantee full separability.

In this paper we have settled the degree conjecture for the separability of multipartite quantum states. Recently, Hildebrand, Mancini  and  Severini have proved that the degree criterion is equivalent to the PPT criterion for separability [17]. However, the importance of degree conjecture ensues from the opportunity it offers to test the strengths and limitations of a nascent approach to the separability problem. We see that this approach has contributed to test the separability of a class of multipartite mixed states (Theorem 5.1) as well as to the efficient factorization of multipartite pure states.\\

\textbf{ACKNOWLEDGMENTS}\\

We thank Dr. Guruprasad Kar and Professor R. Simon for encouragement. ASMH thanks  the Government of Yemen for financial support. We thank Bhalachandra Pujari for his help with LaTex.\\

\begin{verse}

[1] S.L.Braunstein, S. Ghosh,  T. Mansour, S. Severini, R. C. Wilson, Physical Review A \textbf{73},012320 (2006).\\

[2] Ali Saif M. Hassan and Pramod S. Joag , J. Phys. A: Math. Theor. \textbf{40}, 10251 (2007).\\

[3] M. B. Plenio and S. Virmani, Quantum Inf. Comput. \textbf{7}, 1 (2007).\\

[4] K. \.Zyczkowski and I. Bengstsson, quant-ph/0606228.\\

[5] A. Peres, Phys. Rev. Lett. \textbf{77}, 1413 (1996).\\

[6] K. Chen and L.-A. Wu, Quantum Inf. Comput. \textbf{3}, 193 (2003), O. Rudolph, Phys. Rev. A \textbf{67}, 032312 (2003),  O. Rudolph, quant-ph/0202121.\\

[7] M., P., R. Horodecki, Phys. Lett. A \textbf{223}, 1 (1996).\\

[8] B. M. Terhal, Phys. Lett. A \textbf{271}, 319 (2000).\\

[9] Julio I. de Vicente, Quantum Inf. Comput. \textbf{7}, 624 (2007).\\

[10] Ali Saif M. Hassan and Pramod S. Joag, quant-ph/0704.3942.\\

[11] O. Guhne, P. Hyllus, O. Gittsovich and J. Eisert, quant-ph/0611282v2.\\

[12] O. Guhne, Phys. Rev. Lett. \textbf{92}, 117903 (2004).\\

[13] S. Braunstein, S. Ghosh, S. Severini,  Ann. of combinatorics, volume \textbf{10}, No. 3, 2006. e-print quant-ph/0406165.\\

[14] W. Imrich and Klavzar, {\it Product Graphs, Structure and Recognition} (With a forward by Peter Winkler, Wiley- Interscience Series in Discrete Mathematics and Optimization, Wiley- Interscience, New York, 2000).\\

[15] Mohar B 1991 {\it The Laplacian spectrum of graphs} (Graph theory combinotorics and Applications vol.II, wiley).\\

[16] R., P., M. and  K. Horodecki, quat-ph/0702225v1.\\
 
[17] R. Hildebrand , S. Mancini  and  S. Severini  arXiv: cs. CC/0607036.\\

\end{verse}

\end{document}